\newcommand{\ptl}{{\partial}}
\newcommand{\px}{{\partial}_{x}}

\newcommand{\ben}{\begin{equation}}
\newcommand{\een}{\end{equation}}
\newcommand{\bea}{\begin{eqnarray}}
\newcommand{\eea}{\end{eqnarray}}
\newcommand{\nn}{\nonumber\\ }

\newcommand{\sq}{star-quantization}

\renewcommand{\*}{\star}

\newcommand{\ep}{{\epsilon}}
\newcommand{\qq}{\qquad\qquad}
\newcommand{\QQ}{\qquad\qquad\qquad\qquad}
\input amssym.def
\def\Z{{\Bbb Z}}   \def\R{{\Bbb R}}

\documentclass[genTeX]{nrc1}

 \volyear{83}{2005} \journal{Can. J. Phys.}
\received{July 31, 2005} \accepted{?????}
\begin{document}

\title{Star-quantization of an infinite wall}
\author[S. Kryukov]{Sergei Kryukov}
\address{Bogoliubov Laboratory for Theoretical Physics, Joint
Institute for Nuclear Research, 141980 Dubna (Moscow Region),
Russia. \email{kryukov@itp.ac.ru}}
\author[M.A. Walton]{Mark A. Walton}
\address{Department of Applied Mathematics, University of Western Ontario,
London, ON  N6A 5B7, Canada; on leave from: Department of Physics,
University of Lethbridge, Lethbridge, AB  T1K 3M4, Canada.
\email{walton@uleth.ca}}
\shortauthor{Kryukov \& Walton}
\maketitle
\begin{abstract} In deformation quantization (a.k.a. the
Wigner-Weyl-Moyal formulation of quantum mechanics), we consider a
single quantum particle moving freely in one dimension, except for
the presence of one infinite potential wall. Dias and Prata
pointed out that, surprisingly, its stationary-state Wigner
function does not obey the naive equation of motion, i.e. the
naive stargenvalue ($\*$-genvalue) equation. We review our recent
work on this problem, that treats the infinite wall as the limit
of a Liouville potential. Also included are some new results: (i)
we show explicitly that the Wigner-Weyl transform of the usual
density matrix is the physical solution, (ii) we prove that an
effective-mass treatment of the problem is equivalent to the
Liouville one, and (iii) we point out that self-adjointness of the
operator Hamiltonian requires a boundary potential, but one
different from that proposed by Dias and Prata.
\end{abstract}
\begin{resume}
French version of abstract (supplied by CJP if necessary)
   \traduit
\end{resume}

\section{Introduction}

Deformation quantization is also known as the Wigner-Weyl-Moyal
formulation of quantum mechanics, and by other names.\footnote{For
an elementary review, see \cite{HWW}, which cites several more
advanced reviews.} We'll refer to it here as star-quantization. In
it, as in classical mechanics, observables are realized as
ordinary functions (and distributions) on phase space. They are
multiplied, however, using a non-commutative, associative product,
known as the star product ($\*$-product).

Star-quantization can be understood as the Wigner-Weyl transform
of the quantum dynamics of the density matrix. On the other hand,
it is an autonomous way of doing quantum mechanics. Quantum
systems should therefore be treatable in it, without reference to
operators and wave functions, or to path integrals, or to any
other quantum formulation. That is, physical systems should be
star-quantizable.

This note is concerned with the star-quantization of a very simple
system, one particle moving freely along the negative $x$-axis, in
the presence of an infinite potential wall at $x=0$. Naive
star-quantization does not work for it \cite{DP}. We would like to
understand why it fails, and how the problem can be fixed.

Our considerations will be restricted to stationary states of the
system. As emphasized in \cite{CFZ}, where several examples were
worked out, such stationary states are described by
time-independent Wigner functions that can be found as solutions
of the so-called $\*$-genvalue equations.

We first introduce the $\*$-genvalue equations in section 2, and
solve them for the Wigner function of a free particle moving on
the entire $x$-axis. In the following section, the original
problem \cite{DP}, occurring when an infinite barrier is present
at $x=0$, is described. In section 4, we review our treatment
\cite{KW} of the star-quantization of an infinite wall as the
limit of an exponential potential. An alternative to the
$\*$-genvalue equation was found that can be treated in the naive
way: it can be solved first, and then the boundary conditions can
be imposed later, to yield the physical Wigner function. In
section 5, we note that the system can be treated using an
effective mass. Interestingly, however, we show that the
corresponding $\*$-genvalue equations are equivalent to those for
the exponential potential. The ad hoc resolution of the problem
proposed by Dias and Prata \cite{DP} (see also \cite{DPi, DPii})
is then described, in section 6. We have found in operator quantum
mechanics a motivation for a result similar in form to, but in
conflict with theirs. This is sketched in the same section 6. In
\cite{KWi}, we plan to work out further consequences in
star-quantization of this observation.

\section{Free particle: $\*$-genvalue equation \& Wigner-Weyl transform}

Consider a single particle moving on the $x$-axis, so that its
phase space is $\R^2$, with coordinates $x,p$. In \sq, the Wigner
function encodes all observable information about the quantum
state. Stationary states are described by time-independent Wigner
functions $\rho(x,p)$, that can be found as solutions to the
so-called $\*$-genvalue equations \ben H\*\rho\ =\ \rho\* H\ =
E\rho\ \ .\label{sve}\een Here, $H=H(x,p)$ is the classical
Hamiltonian, $E$ is the energy, and the Gr\"onewold-Moyal
$\*$-product is defined by \ben f(x,p)\* g(x,p)\ =\
\exp\left[i\hbar
\left(\ptl_x\ptl_{p'}-\ptl_p\ptl_{x'}\right)/2\right]
\,f(x,p)\,g(x',p')\big\vert_{(x',p')=(x,p)}\ .\label{stp}\een Real
solutions of these equations will describe quantum states, but
they will be mixed states, in general. In this work, we will only
consider pure states, whose Wigner functions obey the additional
$\*$-projection condition \ben \rho_i\*\rho_j\ =\
\delta_{ij}\,\rho_j/h\ ,\ \ {\rm or}\ \ \ \rho_\alpha\*\rho_\beta\
\propto\ \delta(\alpha-\beta)\,\rho_\beta/h\ ,\label{spc}\een for
states labelled by discrete parameters $i,j,\ldots$, or for
non-normalizable states labelled by continuous parameters
$\alpha,\beta,\ldots$, respectively.

To illustrate, let us first consider a free particle (see the
Appendix of \cite{KW}). The Hamiltonian is $H=p^2$, if we set
$2m=1$ for simplicity. The $\*$-genvalue equations (\ref{sve})
then reduce to \ben (p-i\hbar\ptl_x/2)^2\rho(x,p)\ =\
k^2\rho(x,p)\ ,\label{svef}\een where we have put $E=k^2$, and we
will work henceforth with $\hbar=1$. The imaginary part of this
equation is $p\,\px\rho\ =\ 0$, so that when $p\not=0$,
$\px\rho=0$, but not when $p=0$. This leads to the ansatz  \ben
\rho(x,p)\ =\ f(p) + \delta(p)\,g(x)\ .\label{ans}\een In the real
part of (\ref{svef}), $(p^2-\px^2/4 )\,\rho\ =\ k^2\,\rho$, it
yields \ben (p^2-k^2)\,f(p)\ -\ \delta(p)( k^2+ \px^2/4 )\,g(x)\
=\ 0\ .\label{Rfi}\een Considering $p=0$ gives \ben g(x)\ =\
b\exp(2ikx)\ +\ b^*\exp(-2ikx)\ , \label{bpm}\een where
$\rho=\rho^*$ has been imposed. Then (\ref{Rfi}) reduces to
$(p^2-k^2)\,f(p)=0$, solved by \ben f(p)\ =\ a_+\delta(p-k)\ +\
a_-\delta(p+k)\ ,\label{apm}\een with $a_\pm$ arbitrary real
constants. The general result is therefore \ben \rho\ =\
\delta(p)\,\big\{\, b\exp(2ikx)\ +\ b^*\exp(-2ikx)\,\big\}\ +\
a_+\delta(p-k)\  +\ a_-\delta(p+k)\ .\label{rhog}\een

To restrict to pure-state Wigner functions, we impose $\rho\*\rho\
\propto\ \delta(0)\,\rho$, a special case of (\ref{spc}). We find
the constraint \ben |b|^2\ =\ a_+a_-\ \ \Rightarrow\ \ b\ =\
\sqrt{a_+a_-}\, e^{i\phi}\ ,\ \phi\in{\R}\, . \label{purecon}\een
The general pure-state solution to the free-particle $\*$-genvalue
equation (\ref{svef}) is therefore \ben \rho\ =\ a_+\delta(p-k)\ \
+\ \ a_-\delta(p+k)\
 +\ \
2\sqrt{a_+a_-}\,\,\delta(p)\,\cos\big(2kx+\phi\big)\
.\label{rhopure}\een

On the other hand, the Wigner function can be calculated as the
(normalized) Wigner-Weyl transform of the density matrix
$|\psi\rangle\langle\psi|$, built from the known wave functions:
 \ben \rho[\psi]\ :=\ \frac{1}{\pi}\,\int^\infty_{-\infty} dy\,
e^{-2ipy}\,\psi(x+y)\,\psi^*(x-y)\ .\label{rhop}\een We denote it
as $\rho[\psi]$ here to emphasize that it is calculated from known
input wave functions, and so is not a result of working within the
star-quantization formulation. With the pure-state wave function
$\psi\ =\ \alpha_+e^{ikx}\ +\ \alpha_-e^{-ikx}$, eqn. (\ref{rhop})
yields \ben \rho[\psi]\ =\ |\alpha_+|^2 \delta(p-k)\ +\
|\alpha_-|^2 \delta(p+k) \ +\ \delta(p)\big\{
\alpha_+^*\alpha_-e^{-2ikx}\ +\ \alpha_+\alpha_-^*e^{2ikx} \big\}\
.\label{rprp}\een Comparing (\ref{rprp}) with (\ref{rhopure})
reveals a one-to-one correspondence. All is well for the free
particle.

\section{Infinite wall: $\*$-genvalue equation vs Wigner-Weyl transform}

Now consider the Hamiltonian $H=p^2+V(x)$, with potential energy
\ben V(x)\ =\ \left\{ \matrix{0\, , & x<0\, ; \cr \infty\, , &
x>0\, .}\right. \label{Vinf}\een We first work out the Wigner-Weyl
transform of the known Schr\"odinger wave functions
 \ben \psi(x)=\theta(-x)\,\sin(kx)\ .\label{psiL}\een Here
 $\theta(x)$ denotes the Heaviside step function. From
 (\ref{rhop}) we find \ben \rho[\psi]\ \propto\
\theta(-x)\,\bar\rho(x,p)\ ,\ \ \label{rthrb}\een
 with \ben \bar\rho(x,p)\ =\
 \frac{\sin[2x(p+k)]}{2(p+k)}\ +\
 \frac{\sin[2x(p-k)]}{2(p-k)} \ -\
 2\cos(2xk)\,\frac{\sin(2xp)}{2p}\ .\label{rbwi}\een
This can be obtained from a special case of the free-particle
Wigner function (\ref{rhopure}) by the replacement
$\delta(p-p_0)\rightarrow \sin[2x(p-p_0)]/[2(p-p_0)]$. The latter
vanishes at $x=0$, so that $\bar\rho(0,p)=0$, and becomes the
former when $x\to -\infty$, so that a free-particle result is
found far from the infinite wall.

We will show in the following section that (\ref{rthrb},
\ref{rbwi}) describe the physical Wigner function, the one that
should be found in the star-quantization of an infinite wall.

To solve the $\*$-genvalue equations (\ref{sve}) with the
Hamiltonian determined by (\ref{Vinf}), we can try to follow the
Schr\"odinger treatment of this system, by restricting to $x<0$,
and then imposing the boundary condition $\rho(0,p)=0$. For $x<0$,
the $\*$-genvalue equation is that of a free particle, however,
eqn. (\ref{svef}). The problem is that the Wigner function given
by (\ref{rthrb}, \ref{rbwi}) does not satisfy this equation.

\section{Infinite barrier as the limit of an exponential
potential}

To study this problem, we treated the Hamiltonian of the previous
section as the $\alpha\to\infty$ limit of \ben
H_\alpha=p^2+e^{2\alpha x}\ . \label{Halpha} \een
Star-quantization of this system had already been carried out at
$\alpha=1$, in \cite{CFZ}. The $\alpha$-dependence was easily
reinstated. The physical Wigner function is given by (cf. eqn.
(98) of \cite{CFZ})\bea \rho_\alpha&\propto\ \ \   \int_C ds\,
\left[\frac{e^{4\alpha x}}{(2\alpha)^4} \right]^s\,
\Gamma\Big(\frac{i}{2\alpha}(p-k)-s\Big)\,
\Gamma\Big(\frac{i}{2\alpha}(p+k)-s\Big)\, \nn &\QQ\ \ \ \times\,
\Gamma\Big(\frac{-i}{2\alpha}(p-k)-s\Big)\,
\Gamma\Big(\frac{-i}{2\alpha}(p+k)-s\Big)\ . \label{rhoal}\eea The
contour $C$ in the $s$ plane runs from $-i\infty$ to $+i\infty$,
just to the left of the four poles on the imaginary $s$ axis at
$\pm\frac{i}{2\alpha}(p\pm' k)$. The right-hand side of
(\ref{rhoal}) is the Mellin-Barnes type integral definition of the
Meijer G-function (see \S5.3. of \cite{E}), so \ben
\rho_\alpha\propto\ G^{40}_{04}\left( \frac{e^{4\alpha
x}}{(2\alpha)^4} \bigg| \frac{i(p-k)}{2\alpha},
\frac{i(p+k)}{2\alpha}, \frac{-i(p-k)}{2\alpha},
\frac{-i(p+k)}{2\alpha} \right)\ . \label{rhoalG}\een

First, consider the $\alpha\to\infty$ limit of the Wigner
function. For $x>0$, $w=e^{4\alpha x}/(2\alpha)^4\to\infty$, but
$G^{40}_{04}(w|\cdot)\to 0$ exponentially, according to the
asymptotics described in \S5.4.1 of \cite{E}.

For $x<0$, the contour $C$ can be closed such that all 4 poles on
the imaginary $s$-axis, at $s\ =\ {i}(\pm p\pm' k)/{2\alpha}$, are
surrounded. First note that \ben \lim_{\alpha\to\infty}\,
(2\alpha)^{-2i(\pm p\pm' k)/\alpha}\ =\ 1\ .\label{lone}\een We
can now apply the residue theorem and use
$\Gamma(z)=\Gamma(1+z)/z$, to find $\rho_\alpha(x,p)$ goes as \bea
&\ e^{2ix(p-k)}/ {(-2ik)2(ip-ik)(2ip)}\ +\
e^{2ix(p+k)}/{(2ik)(2ip)2(ip+ik)}\ \qq\nn &+\
e^{-2ix(p-k)}/{2(-ip+ik)(-2ip)(2ik)}\ +\
e^{-2ix(p+k)}/{(-2ip)2(-ip-ik)(-2ik)}\ ,\label{lalr}\eea as
$\alpha\to\infty$, up to a factor depending on $\alpha$, but
independent of $x.$ But this is proportional to \bea &\
e^{2ix(p-k)}\left(\frac1{kp}-\frac1{k(p-k)}\right)\ +\
e^{2ix(p+k)}\left(\frac1{kp}-\frac1{k(p+k)}\right)\qq\nn &-\
e^{-2ix(p-k)}\left(\frac1{kp}-\frac1{k(p-k)}\right)\ -\
e^{-2ix(p+k)}\left(\frac1{kp}-\frac1{k(p+k)}\right)\
.\label{lalr}\eea Comparing with (\ref{rbwi}), we see that for
$x<0$, we have shown that
$\lim_{\alpha\to\infty}\rho_\alpha\,\propto\, \bar\rho$.

Combining the results for $x<0$ and $x>0$, we have that
$\lim_{\alpha\to\infty}\rho_\alpha\,\propto\,
\theta(-x)\,\bar\rho$. That is, the limit of the Wigner function
for the exponential potential is $\theta(-x)\,\bar\rho$, showing
that (\ref{rthrb}) and (\ref{rbwi}) describe the physical Wigner
function for the infinite potential wall. The authors of
\cite{DP}, e.g., made this assumption.

Our goal is to find a dynamical equation that this physical Wigner
function satisfies, instead of the $\*$-genvalue equation, which
it does not. The new equation \cite{KW} can then be used to solve
for the physical Wigner function, i.e., the new equation can be
used to carry out a true star-quantization of the infinite
potential wall.

Following \cite{CFZ}, we see that (\ref{Halpha}) yields a
$\*$-genvalue equation $H_\alpha\star\rho_\alpha\ =\
k^2\,\rho_\alpha$ with imaginary and real parts \bea &e^{-2\alpha
x}\,\partial_{x}\rho_\alpha(x,p)\ =-\ \frac{i}{2p}
\left[\rho_\alpha(x,p+i\alpha)-\rho_\alpha(x,p-i\alpha)\right ]\ ,
\label{IHr}\cr &e^{ -2\alpha x}\left
(p^2-k^2-\frac{1}{4}\,\partial_{x}^{2}\right )\rho_\alpha(x,p)\ +\
\frac{1}{2}\left
[\rho_\alpha(x,p+i\alpha)+\rho_\alpha(x,p-i\alpha)\right ]\ =\ 0\
,\ \label{IHr}\eea respectively. These two equations can be
recombined into a difference equation:\footnote{At $\alpha=1$,
this is a slight correction of eqn. (104) of \cite{CFZ}.} \bea 0\
=\ &(p^2-k^2)\rho_\alpha(x,p) + \frac{1}{p}\left (\frac{e^{2\alpha
x}}{4}\right )^2
\left[\frac{\rho_\alpha(x,p+2i\alpha)-\rho_\alpha(x,p)}{p+i\alpha}
+\frac{\rho_\alpha(x,p-2i\alpha)-\rho_\alpha(x,p)}{p-i\alpha}\right
]\nn
 & -\frac{i e^{ 2\alpha
x}}{4p}\left[\rho_\alpha(x,p+i\alpha)-\rho_\alpha(x,p-i\alpha)\right]
+ \frac{ e^ {2 \alpha
x}}{2}\left[\rho_\alpha(x,p+i\alpha)+\rho_\alpha(x,p-i\alpha)\right]\
. \label{difff} \eea In \cite{KW}, we showed that $\rho_\alpha(x,p
\pm i\alpha),\ \rho_\alpha(x,p\pm 2i\alpha)$ could be traded for
$\partial_x^n\rho_\alpha(x,p),\ n=1,2,3,4$, to arrive at a
differential equation, with a simple $\alpha\to\infty$ limit, \ben
\partial_{x}^4 \rho(x,p)/16\ +\
(p^2+k^2)\,\partial_{x}^2\, \rho(x,p)/2\ +\
(p^4-2k^2p+k^4)\,\rho(x,p)\ =\ 0\ ,\ \label{newiwall}\een valid
for $x< 0$. Recalling that $E=:k^2$, the new equation can be
rewritten as \ben (p^2-E)\star\rho\*(p^2-E)\ =\ 0\
.\label{ststnew}\een

It is easily verified that $\bar\rho$ of (\ref{rbwi}) satisfies
the new equation. In \cite{KW} it was also shown that the new
equation applies to other systems built from infinite walls or
wells: the infinite square well and the delta-function well.
Furthermore, it was shown how an additional, non-singular
potential could be included in the framework.

The proposal made in \cite{KW} was to star-quantize the
infinite-wall system by solving the new equation, and then
imposing the boundary conditions, much as for solutions of the
Schr\"odinger equation. Some progress in this direction was made
in \cite{DPii}.

\section{Effective mass treatment}

Let us mention another possible approach. It is clear that the
free Hamiltonian $H=p^2$ encodes nothing of the dynamics of the
infinite potential wall at $x=0$. In the Schr\"odinger treatment,
imposing the boundary conditions suffices to yield the required
physics, but not in star-quantization. Can the dynamics of the
wall be described by a modified Hamiltonian?

One could try to incorporate the dynamics in the kinematics, by
assigning to the particle an effective mass that blows up for
$x>0$. In the spirit of the last section, such a Hamiltonian can
also be considered as a limit: \ben H_{\rm eff}'\ :=\
\theta(-x)\,p^2\ =\ \lim_{\alpha\to\infty}\, H_{{\rm
eff}\hskip-1pt,\,\alpha}'\ :=\ \lim_{\alpha\to\infty}\,
\left(1+e^{2\alpha x} \right)^{-1}\,p^2\ \ .\label{effa}\een

We will now show that \ben H_{{\rm
eff}\hskip-1pt,\,\alpha}'(x',p)\* \rho'_{\rm eff}(x',p)\ =\
k^2\,\rho'_{\rm eff}(x',p)\ \label{Hasve}\een is equivalent to the
$\*$-genvalue equation with Hamiltonian $H_\alpha$ of
(\ref{Halpha}), already considered.

Using (\ref{stp}), one can show that (\ref{Hasve}) reduces to \ben
(p-i\partial_{x'}/2)^2\,\rho_{{\rm
eff}\hskip-1pt,\,\alpha}'(x',p)\ =\ k^2\,(1+e^{2\alpha
x'+i\alpha\partial_p})\,\rho'_{{\rm
eff}\hskip-1pt,\,\alpha}(x',p)\ .\label{Hasx}\een Writing $ x'\ =\
x\ +\ \log(-k^2)/(2\alpha)\ \label{xxp}$ yields \ben
(p-i\partial_x/2)^2\,\rho_\alpha(x,p)\ =\ (k^2-e^{2\alpha
x+i\alpha\partial_p})\,\rho_\alpha(x,p)\ ,\label{Hasxp}\een
however, where we put $\rho_\alpha(x,p):=\rho'_{{\rm
eff}\hskip-1pt,\,\alpha}(x',p)$. The transformed equation is just
$H_\alpha(x,p)\*\rho_\alpha(x,p)\,= \,k^2\rho_\alpha(x,p)$, the
$\*$-genvalue equation for (\ref{Halpha}). Since $x'-x\to 0$ in
the $\alpha\to\infty$ limit, (\ref{Hasve}) must lead to the same
eqn. (\ref{ststnew}) we already found above.

\section{Self-adjoint extensions and boundary
potentials}

Let us reconsider the operator treatment of the system, to try to
understand why pure star-quantization fails. Perhaps the problem
will turn out to be the Wigner-Weyl transform of one already
present in the operator formulation.

A self-adjoint Hamiltonian $\hat H$ will have a spectral
decomposition, $\hat H \,=\, \sum_{E'} |E'\rangle\langle E'|$,
roughly speaking. Then for a pure-state density matrix $\hat\rho
\,=\, |E\rangle\langle E|$ built from the stationary energy
eigenstate $|E\rangle$, we necessarily have $\hat H\hat\rho \,=\,
\hat\rho\hat H \,=\, E\hat\rho$. If the Wigner-Weyl transform
$\cal W$ works, these become the $\*$-genvalue equations
(\ref{sve}).

The key realization is that the free Hamiltonian operator $\hat
H:=\hat p^2$ is not self-adjoint on the negative $x$-axis,  even
if it is Hermitian (see \cite{ACP}, e.g., and references therein).
We believe that this is the root cause of at least part of the
problem encountered with naive star-quantization of the infinite
wall.

The free Hamiltonian $\hat H$ does have self-adjoint extensions,
however, as can be determined by calculating its von Neumann
deficiency indices. The extensions amount to including a point
interaction at $x=0$. The role of the point interaction is to
enforce a (Robin) boundary condition on the Schr\"odinger wave
function, involving a real length $L$: \ben \psi(0)\ +\ L\psi'(0)\
=\ 0\ \ \ \ (\,\psi':=d\psi/dx\,)\ .\label{labc}\een It is easy to
see that such a potential takes the form $V_L(x) = \delta(x)\,-\,
L\delta'(x)$.

A specific interaction of this type was prescribed in an ad hoc
way by Dias and Prata, for the boundary condition $\psi(0)=0$. It
is interesting to note, however, that their prescription was for a
boundary potential proportional to $\delta_-'(x)$, a regularized
version of $\delta'(x)$. Apparently, this contradicts the standard
operator treatment, since the Dirichlet boundary condition
corresponds to $L=0$ in (\ref{labc}), which relates to $V_{L=0}(x)
=\delta(x)$.

Clearly, the situation needs clarification. We hope to report
further progress on this problem in \cite{KWi}.

\vskip.5cm\noindent{\bf Acknowledgments}\hfill\break MW thanks the
Applied Math Dept. of the University of Western Ontario for its
kind hospitality. His research was supported in part by  a
Discovery Grant from NSERC of Canada.

\end{document}